\documentclass{PoS}

\usepackage{amsmath,amssymb}
\usepackage{epsfig}

\newcommand{\psibar}{\overline{\psi}}

\newcommand{\zetabar}{\overline{\zeta}}

\newcommand{\gmu}{\gamma_\mu}
\newcommand{\gnu}{\gamma_\nu}
\newcommand{\go}{\gamma_0}

\newcommand{\gfive}{\gamma_5}
\newcommand{\dmu}{\partial_\mu}
\newcommand{\md}{\widetilde{\partial}}

\newcommand{\e}{\mathrm{e}}
\newcommand{\rd}{\mathrm{d}}

\newcommand{\rO}{\mathrm{O}}
\newcommand{\<}{\langle}
\newcommand{\z}{\rangle}

\newcommand{\li}{\left}
\newcommand{\re}{\right}
\newcommand{\ren}[1]{\li(#1\re)_\mathrm{R}}
\newcommand{\ev}[1]{\bigl\<#1\bigr\z}

\PoS{PoS(LAT2005)266}

\title{Perturbative renormalisation of the chiral Gross-Neveu model\footnote{ This work was supported by the Deutsche Forschungsgemeinschaft in form of Graduiertenkolleg GK 271 and Sonderforschungsbereich SFB TR 09.}}

\ShortTitle{Perturbative renormalisation of the chiral Gross-Neveu model }
\author{\speaker{Bj\"orn Leder}\\
        DESY Zeuthen\\
        E-mail: \email{bjoern.leder@desy.de}}

\author{Tomasz Korzec\\
        Humboldt University of Berlin\\
        E-mail: \email{korzec@physik.hu-berlin.de}}

\abstract{We study the chiral Gross-Neveu model with Wilson fermions. In the framework of the Schr\"odinger functional we show that in general
not only the bare mass has to be tuned to achieve chiral symmetry in the continuum, but also 
coupling constants multiplying chirally non-invariant four fermion terms.
The strategy for fixing the parameters of the theory is explained in perturbation theory and
the results of a first order calculation are presented. The results are shown to agree with the infinite flavour limit.
\vspace{1cm}
\flushright{DESY 05-187\\HU-EP-05/60\\SFB/CPP-05-59\\}}

\FullConference{XXIIIrd International Symposium on Lattice Field Theory\\
		 25-30 July 2005\\
		 Trinity College, Dublin, Ireland}

\begin{document}

\section{Introduction}

Theories of self-coupled fermions in two dimensions may show interesting features like asymptotic freedom and chiral symmetry \cite{Gross:1974jv}. A model with these properties provides a lab for all sorts of questions 
raised in the context of lattice formulations of chiral fermions in four dimensions. In principle, the universality of different
fermion discretizations can be investigated, new ideas can be tested in a controlled environment or ``established'' concepts can be challenged with high precision Monte Carlo (MC) data.

Many of the perturbative and nonperturbative properties of the continuum chiral Gross-Neveu model (CGN) have been calculated, see e.g. \cite{Forgacs:1991nk} and \cite{Bennett:1999bc}. The properties of chiral symmetry with Wilson fermions have been studied in the limit of an infinite number of flavours \cite{Aoki:1985jj, Izubuchi:1998hy}. In this work we use lattice perturbation theory (PT) to compute the critical mass and other parameters of the lattice action with Wilson fermions.

\section{Self-coupled fermions}

Rather than just stating an action associated with CGN we start with a general action containing all possible terms that self-couple the fermions. Their number will be reduced by symmetries and interdependence. In this way we are sure to include all interaction terms that might be important for renormalisation.

In two dimensions fermion fields $\psibar$, $\psi$ have mass
dimension $1/2$. Therefore the couplings multiplying four fermion interactions (4FI) are dimensionless and relevant. In the continuum euclidean
action for $N$ flavours of Dirac fermions\footnote{If not stated otherwise, repeated small letter indices are summed over.}
\begin{equation}
	S = \int \rd x^2\, \li\{ \psibar_{\alpha i}\li((\gmu)_{\alpha \beta}\dmu+m\delta_{\alpha \beta}\re)\psi_{\beta i} +
  	 C_{\alpha\beta\gamma\delta}^{ijlk}\; \psibar_{\alpha i}\, \psi_{\beta j}\, \psibar_{\gamma k}\, \psi_{\delta l} \re\}
\end{equation}
the 4FI consist of four Dirac spinors with Dirac (Greek letters $\alpha,\beta,\dots$) and flavour (Latin letters $i,j,\dots$) indices contracted by the tensor $C$.

Imposing now a set of symmetries: flavour $\text{U}(N)$ symmetry, Euclidean invariance, parity and using the properties of $\gamma$-matrices in two dimensions ($\gmu\gfive=i\epsilon_{\mu\nu}\gnu$) one finds that $C$ can be written as a superposition of 6 terms characterized by three different Dirac structures and by two different flavour structures:\footnote{The contraction of Dirac and flavour indices is obvious, hence they are suppressed.}
\begin{equation}\label{sixterms}
\begin{array}{ll}
	O_{SS} =(\psibar\psi)^2 				&\quad O'_{SS} =\sum_a(\psibar\lambda^a\psi)^2\\
	O_{PP} =(\psibar i\gfive\psi)^2 		&\quad O'_{PP} =\sum_a(\psibar i\gfive\lambda^a\psi)^2\\
	O_{VV} =\sum_\mu(\psibar\gmu\psi)^2 	&\quad O'_{VV} =\sum_{\mu,a}(\psibar\gmu\lambda^a\psi)^2
\end{array}
\end{equation}
In the flavour non-singlet terms on the right hand side the sum is over the $N^2-1$ generators $\lambda^a$ of $\mathrm{SU}(N)$. But these terms
are not all independent. Fierz transformations (FT) allow to interchange products of Dirac spinor bilinears like the terms in \eqref{sixterms}. In two dimensions the FT yield three equations like
\begin{equation}
	O_{SS} - O_{PP} = -\tfrac{1}{N} O_{VV} - \tfrac{1}{2} O'_{VV}\,.
\end{equation}
Taking these interdependencies into account there are {\it three} independent 4FI and hence {\it three} dimensionless couplings\footnote{For $N=1$ the number of degrees of freedom allow only for one local interaction term. Here we consider $N>1$.}.

Chiral symmetry is not considered at this stage, because we want to study the theory on the lattice with the Wilson discretization, which
breaks chiral symmetry explicitly. One can choose any three terms from \eqref{sixterms}. In view of nonperturbative MC calculations we
choose the three terms with flavour singlet bilinears only. To be specific, on a cubic lattice with extension $T\times L$ we study the theory with the euclidean action
\begin{equation}\label{latticeaction}
	S = a^2\sum_x\,\li\{\psibar(D_W +m_0)\psi-\frac{1}{2}\li(g_S^2\,O_{SS} + g_P^2\, O_{PP} + g_V^2\, O_{VV}\re) \re\}\,,
\end{equation}
where $D_W$ is the Wilson-Dirac operator and $a$ the lattice spacing.

\vspace{-0.2cm}
\section{Chiral Ward-Takahashi identity}
\vspace{-0.2cm}

The PCAC (partially conserved axial current) relation of the theory obtained in the naive continuum limit of \eqref{latticeaction} takes the
usual form, but is modified by a term originating from the scalar and pseudo scalar 4FI
\begin{equation}\label{PCAC}
	\ev{O(y)\,\dmu A_\mu(x) } = \ev{ O(y)\, 2\,m_0\, P(x)} - \ev{O(y)\, (g_S^2-g_P^2)/2\, P(x)S(x)}\,, \quad x\neq y\,,
\end{equation}
where
\begin{equation}
	A_\mu(x) = \psibar(x)\gmu\gfive\psi(x)\,,\quad P(x) = \psibar(x)\gfive\psi(x)\,,\quad S(x) = \psibar(x)\psi(x)\,.
\end{equation}
The vector 4FI is chirally invariant on its own. Note that an $\mathrm{U}(1)$ chiral transformation was used to derive this and therefore only the flavour singlet current and densities appear. Imposing chiral symmetry in the continuum theory the mass has to vanish $m_0=0$
and the couplings of the scalar and the pseudo scalar 4FI have to be equal $g_P^2 = g_S^2$. The PCAC relation then turns into the
conservation of the axial current. The resulting action with the interaction $-\tfrac{1}{2}\li(g_S^2\,(O_{SS} + O_{PP}) + g_V^2\, O_{VV}\re)$ is
what in the literature is called the CGN \footnote{In most large $N$ studies the vector term is not present.}.

However, at finite lattice spacing chiral symmetry is explicitly broken by the Wilson term, the mass is not protected against additive renormalisation and $g_P^2 = g_S^2$ is no longer protected. The right hand side of \eqref{PCAC} becomes a mess due to operator mixing. To achieve chiral symmetry in the 
continuum limit means for properly normalized fields and axial current
\begin{equation}\label{dArenOa}
	\ev{O(y)\,\md_\mu \ren{A_\mu}(x) } = \rO(a)\,,
\end{equation}
where $\md_\mu$ is a symmetric lattice difference operator. That is in an asymptotic expansion of correlation functions $\ev{O(y)\,\md_\mu \ren{A_\mu}(x) }$ in the lattice spacing the zeroth order vanishes.
This condition is used to fix the parameters of the action \eqref{latticeaction} in MC simulations \cite{Korzec:pos} and in PT.

\vspace{-0.1cm}
\section{Correlation functions}
\vspace{-0.1cm}

We define correlation functions in the Sch\"odinger functional (SF) setup \cite{Sint:1993un} in order to utilize~\eqref{dArenOa} \cite{Luscher:1996vw}.
In this finite size scheme the fermion fields obey Dirichlet boundary conditions in time and generalized periodic boundary conditions with a phase shift parametrized by an angle $\theta$ in space
\begin{equation}\label{BCspace}
	\psi(x+L)=\e^{i\theta}\psi(x)\,, \quad \psibar(x+L)=\e^{-i\theta}\psibar(x)\,, \quad \theta\in[0,2\pi)\,.
\end{equation}
The Dirichlet boundary conditions in time may give rise to new counter terms defined on the boundary, in the sense that the associated bare coefficients are needed to absorb infinites on the way to the continuum limit. Relevant operators or composite fields living at the boundary are those which have mass dimension one or less. Thus objects like $\psibar \Gamma \psi$ with $\Gamma=\{1,\go,\gamma_1,\gfive\}$ can appear. But all these terms either vanish because they violate parity or for the boundary fields that we employ. Thus no additional boundary counter terms have to be considered.

The correlation functions
\begin{align}
	f_X(x_0) = -\frac{a^2}{2N_f}\,\sum_{y_1,z_1}\;
  	 	\ev{\;\psibar(x)\,\Gamma_X\,\psi(x)\; \zetabar(y_1)\,\gfive\,\zeta(z_1)\;}\,,
  	 	\quad \Gamma_A=\go\gfive\,,\quad \Gamma_P=\gfive\,, \label{fX}
\end{align}
to be considered are correlators of a zero momentum pseudo scalar boundary state built from the boundary fields $\zeta$, $\zetabar$ and insertions of the time component of the axial current ($f_A$) and the pseudo scalar density ($f_P$) respectively.
For details on the definition of the generating functional and its perturbative expansion we refer the reader to a forthcoming publication \cite{Leder:inprep}.

\section{First order renormalization}

In order to calculate the parameters that ensure \eqref{dArenOa} in first order of PT, the correlation function \eqref{fX} has to be expanded to this order
\begin{equation}\label{fXepansion}
	f_X(x_0) = f_X^{(0)}(x_0) + g_S^2\, f_X^{(1,S)}(x_0) + g_P^2\, f_X^{(1,P)}(x_0) + g_V^2\, f_X^{(1,V)}(x_0) + \rO(g^4)\,,
\end{equation}
where we assume all three couplings to be small and comparable to some small $g^2$, i.e. 
\begin{equation}\label{coeffs}
	g_i^2 = c_i\, g^2\,, \quad \text{and} \quad c_i = \rO(1)\,, \quad i=S,P,V\,.
\end{equation}
\begin{figure}
	\centering
	\epsfig{file=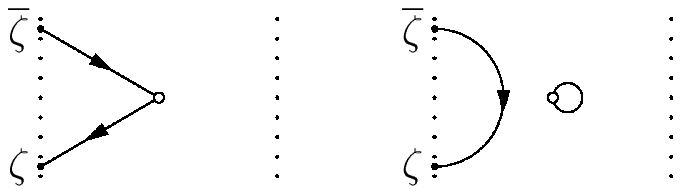,scale=0.8}
	\caption{Tree level diagrams for $f_X$.}
	\label{fig:TreeLevelDiagramsForFX}
\end{figure}
\begin{figure}
	\centering
	\epsfig{file=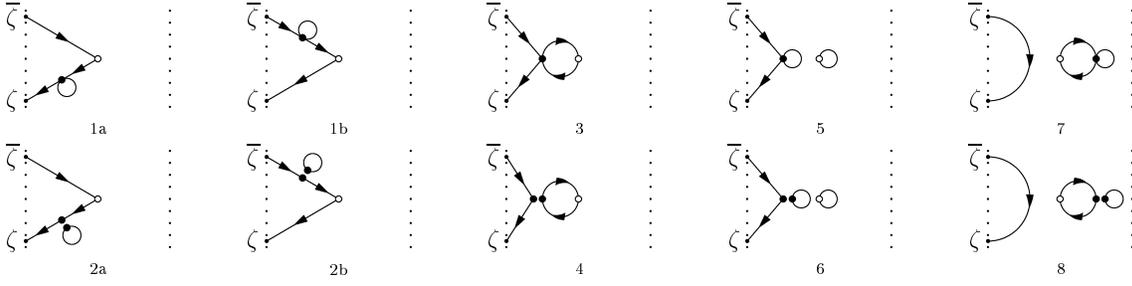,scale=0.8}
	\caption{First order diagrams for $f_X$.}
	\label{fig:FirstOrderDiagramsForFX}
\end{figure}
The tree level amplitude $f_X^{(0)}$ is the sum of the two diagrams sketched in Fig. \ref{fig:TreeLevelDiagramsForFX}. In these diagrams the dotted lines represent the time slices $x_0=0$ and $x_0=T$. Plain lines are used for the fermion propagator and the small open circle in the middle symbolises the insertion.

The first order amplitude $f_X^{(1)}=c_S\,f_X^{(1,S)} + c_P\, f_X^{(1,P)}(x_0) + c_V\, f_X^{(1,V)}$ is a sum of three contributions from the 
three 4FI. Each such contribution is itself a sum of 10 diagrams. Due to the $\gfive$-hermiticity of the Wilson Dirac operator, and thus the fermion propagator, the diagrams 1a and 1b and also 2a and 2b in Fig. \ref{fig:FirstOrderDiagramsForFX} are equal. In these diagrams the small filled circles (dots) represent the 4FI. The diagrams in the lower row of Fig. \ref{fig:FirstOrderDiagramsForFX} are very similar to the respective ones in the upper row but have two separated dots. Because the 4FI consists of two Dirac spinor bilinears, there are always two similar but distinct contractions: either the flavour and Dirac indices of the four spinors are contracted collectively or the two bilinears are contracted separately. One single dot refers to the former and two separated dots to the latter possibility.

We define now the renormalised correlation function $\ren{f_A(x_0)}$ and $h_A(\theta,x_0/L)$ as dimensionless quantities on a lattice with fixed ratio $T/L$
\begin{equation}\label{hA}
	\ren{f_A(x_0)} = Z_A\, Z_\zeta^2\; f_A(x_0) \quad \text{and} \quad 
					h_A(\theta,x_0/L) = L\md_0\,\ren{f_A(x_0)}\,,
\end{equation}
where we introduced normalization factors for the axial current and the boundary fields
\begin{align}
	Z_A = 1 +  g^2\, Z_A^{(1)} + \rO(g^4)\,,\quad	Z_\zeta = 1 +  g^2\, Z_\zeta^{(1)} + \rO(g^4)\,.
\end{align}
Eq. \eqref{dArenOa} means in terms of these functions
\begin{equation}\label{hAcond}
	h_A(\theta,x_0/L) = \rO(a/L)\,,\quad  \forall\,\theta\,,x_0/L\,,\quad \text{at} \quad am_0=am_c\,.
\end{equation}
Employing this condition at tree level one finds that the critical mass $m_c$ vanishes, i.e. \mbox{$am_c=\rO(g^2)$}. Here we determine the first order coefficient in
\begin{equation}\label{amc}
	am_c = am_c^{(1)}\,g^2 + \rO(g^4)\,.
\end{equation}
Using expansions \eqref{fXepansion} and \eqref{amc} in \eqref{hA} gives
\begin{equation}
	h_A = h_0 + g^2 \Big\{ h_1 + am_c^{(1)}\, h_2 + \li(Z_A^{(1)} + 2 Z_\zeta^{(1)}\re)\, h_0 \Big\} + \rO(g^4)\,,
\end{equation}
with
\begin{equation}
	h_0=L\md_0 f_A^{(0)}(x_0)\,, 
		\quad h_1=L\md_0 f_A^{(1)}(x_0)\,,
		 \quad h_2=\frac{\partial}{\partial am_0}h_0\,,
\end{equation}
all defined at $am_0=0$. The tree level amplitude $h_0$ turns out to be $\rO(a^2)$, whereas $h_2$ diverges linearly in $a$. Therefore \eqref{hAcond} amounts to
\begin{equation}\label{final}
	am_c^{(1)} = -h_1/h_2 + \rO(a^2)\,.
\end{equation}
Since the amplitude $h_1$, as $f_A^{(1)}$, is linear in the coefficients $c_i$ introduced in \eqref{coeffs} the finite and the $\rO(a)$
part of $-h_1/h_2$ yield two linear equations for each pair of $\theta$ and $x_0/L$. But the limit $a\to0$, defining the critical mass, turns out to be independent of $\theta$ and $x_0/L$ giving a single relation between the critical mass and the coefficients. The $\rO(a)$ part has to vanish, thus demanding $c_S=c_P$.
In terms of the coupling constants the result is
\begin{align}
    am_c & = -0.7698004(1) \times \li((2N-1)/2\,g_S^2 + g_P^2/2- g_V^2\re) + \rO(g^4)\,,\label{amc1}\\
    g_P^2/g_S^2 & = 1 + \rO(g^2)\,.\label{gP2}
\end{align}
To compare with the large $N$ result of \cite{Aoki:1985jj} we rescale $g_i^2\to g_i^2/N$ and take the $N\to\infty$ limit in \eqref{amc1}
\begin{equation}\label{largeN}
	am_c = -0.7698004(1) \times g_S^2\,.
\end{equation}
This is the full large $N$ result. The authors neglect the vector 4FI from the beginning, but their result is not changed if it is taken into account. In the large $N$ limit $g_P^2$ is a nontrivial function of $g_S^2$ which is reproduced by our result at first order.

\vspace{-0.1cm}
\section{Conclusion and Outlook}
\vspace{-0.1cm}

Self-coupled fermions in two dimensions with $\text{U}(N)$ flavour and continuous chiral $\text{U}(1)$ symmetry are asymptotically free in the
large $N$ limit and in perturbation theory. We establish this theory on the lattice with the Wilson discretization. In general there
are three dimensionless coupling constants and the bare mass parameter due to the explicitly broken chiral symmetry. Continuum Ward-Takahashi identities are used to enforce chiral symmetry in the continuum limit of the lattice theory fixing the parameters of the action. At first order in perturbation theory, when all couplings are small, we determine the critical mass and the ratio of the pseudo scalar and the scalar coupling. The vector coupling is not fixed at this order. It is not needed for the renormalisation of the ratio. It appears in the expansion of the critical mass which is not seen at large $N$. Nevertheless our result contains the full large $N$ critical mass and also the ratio of the couplings is correctly reproduced. 

A next to leading order calculation is almost finished and covered in a forthcoming publication \cite{Leder:inprep}. With this result we will be in the position to calculate a renormalised coupling and thus a step scaling function in the controlled environment of Wilson fermions. Then a direct comparison with other fermion actions like Ginsparg-Wilson and staggered is possible and will be done.

\paragraph{Acknowledgment}

We would like to thank Rainer Sommer and Francesco Knechtli for helpful discussions and Ulli Wolff for reading the manuscript.

\end{document}